\documentclass[aps,prb,twocolumn,superscriptaddress]{revtex4-1}
\usepackage{amssymb}
\usepackage{physics}
\usepackage{graphicx}
\usepackage{amsmath}
\usepackage{amsthm}
\usepackage{amsfonts}
\usepackage[T1]{fontenc}
\usepackage{array}
\usepackage{multirow}
\usepackage {afterpage}
\usepackage{color}
\usepackage{esint}
\usepackage{bm}
\usepackage{color}
\usepackage{bbm}
\usepackage{hyperref}
\usepackage{babel}
\usepackage{titlesec}
\usepackage{MnSymbol}

\newcommand{\red}[1]{\textcolor{black}{#1}}

\begin{document}
	\title{Excitonic phases in a spatially separated electron-hole ladder model}
	\author{DinhDuy Vu}
	\author{Sankar Das Sarma}
	\affiliation{Condensed Matter Theory Center and Joint Quantum Institute, Department of Physics, University of Maryland, College Park, Maryland 20742, USA}
	
	\begin{abstract}
		We obtain the numerical ground state of a one-dimensional ladder model with the upper and lower chains occupied by spatially-separated electrons and holes, respectively. Under charge neutrality, we find that the excitonic bound states are always formed, i.e., no finite regime of decoupled electron and hole plasma exists at zero temperature. The system either behaves like a bosonic liquid or a bosonic crystal depending on the inter-chain attractive and intra-chain repulsive interaction strengths. We also provide the detailed excitonic phase diagrams in the intra- and inter-chain interaction parameters, with and without disorder. We also comment on the corresponding two-dimensional electron-hole bilayer exciton condensation.
	\end{abstract}
	\maketitle
	
	\section{Introduction}
	
	 Electrons and holes can form bound states through the attractive Coulomb interaction, called excitons. At sufficiently low temperatures, excitons can condense due to their bosonic nature \cite{Keldysh1965, Kozlov1965, Keldysh1968, Comte1982, Nozieres1982}. One exciting experimental possibility is the 2D electron-hole bilayer, where the attraction between the electrons in one layer and the holes in the other layer should lead to interlayer coherent excitonic bosonic condensation \cite{Lozovik1975, Shevchenko1976, Zhu1995, Littlewood1996, Zhu1996}. Recently, transport evidence has been reported for bilayer exciton condensation in several different layer systems \cite{Davis2023, Wang2019, Ma2021}. There is also extensive experimental literature on the closely related phenomenon of spontaneous interlayer coherence in bilayer quantum Hall systems with a total filling of unity, where the electron-hole transformation in a filled Landau level produces an effective exciton condensate \cite{Eisenstein2014, Einstein2019, Liu2022}. In spite of very extensive theoretical literature on the subject, the central conceptual issue of the $T=0$ ground-state quantum phase diagram of the electron-hole bilayers remains problematic since even the basic question of the allowed $T=0$ quantum phases remain unknown and controversial. Many publications claim uncritically that the $T=0$ phase contains the unpaired electron-hole liquid as a possible ground state with a (Mott-like) quantum phase transition from the bosonic exciton liquid to the fermionic electron-hole liquid at weak coupling -- see the discussion and citations in \cite{Wu2015}. We believe this claim to be incorrect, and there is no ground state transition to an electron-hole liquid in bilayer electron-hole systems (or quantum Hall bilayers at a filling factor of unity).
	
	In this work, we theoretically investigate, using the Density matrix renormalization group (DMRG), a two-chain 1D analog of 2D bilayers - a ladder model with two oppositely charged spatially-separated 1D chains. Our system is controlled by two parameters, the intra-chain repulsive interaction $U_1$ and inter-chain attractive interaction $U_2$. We are particularly interested in the $(U_1, U_2)$ phase diagram and the important issue of how many phases it can have. To answer this question, we numerically obtain the ground state using the DMRG method implemented by the ITensor package \cite{ITensor}, which is essentially an exact technique for our purpose. We note that the problem of two coupled 1D chains can also be studied from the bosonization perspective \cite{Orignac1997,Giamarchi2007,Furuya2015,Chou2023} which is useful to describe the thermodynamic phases.  Nevertheless, our non-perturbative DMRG approach is preferred to describe the quantitative excitonic phase diagram over a wide range of parameters even though the finite sizes prevent obtaining the critical phase boundary exactly. The reason is that our system has parabolic dispersion and the interaction strength can be much larger than the hopping strength, which greatly complicates the implementation of the bosonization method.
	
	Intuitively, one may hypothesize a plasma phase when the electrons and holes are concentrated densely enough in their respective channels and the effective $U_2$ is small. In the other limit of large $U_2$, the electron-hole interaction should prevail, making the elementary excitations primarily excitons. We find that there is no such electron-hole liquid phase, and excitons are always favored in the ground state for any non-zero attractive interaction, with the excitons at small $U_2$ having a large size and coherent only over a short distance. Additionally, we observe the crystallization of the bosonic liquid for large $U_1$, resulting in a phase diagram only having two phases: an excitonic bosonic liquid and a crystal. We also study the robustness of this phase diagram against disorder and temperature.
	
	\begin{figure*}
		\centering
		\includegraphics[width=\textwidth]{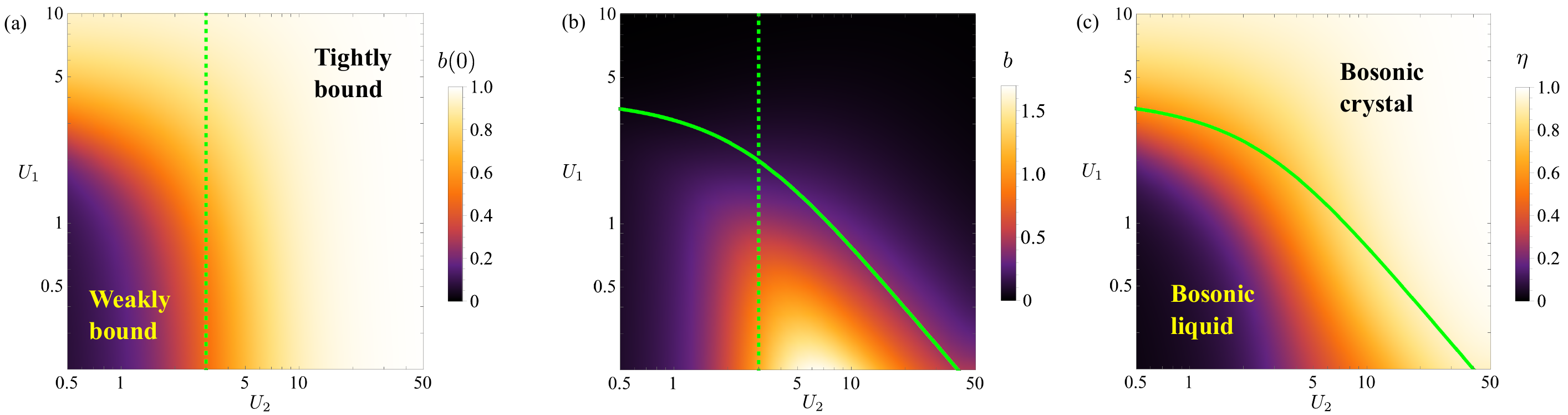}
		\caption{($U_1,U_2$) phase diagram for a 14-rung ladder system. (a) Onsite bosonic correlation $b(0)$. (b) Long-range bosonic correlation $b=\sum_{r=1}^{L/2}b(r)$. (c) Crystalline order parameter from the bosonic density-density correlation. The dashed line given by $U_2=3$ separates the weakly bound $(b(0)<0.5)$ and tightly bound $(b(0)>0.5)$ exciton regimes at weak $U_1$. The solid green line given by $\sqrt{16+U_2^2}-U_2=U_1$ roughly separates the bosonic crystal and bosonic liquid.\label{fig1}}
	\end{figure*}		
	
	\section{Model Hamiltonian}
	The ladder model is defined by
	\begin{equation}\label{model}
	\begin{split}
	H_0 & = -\sum_i (c_{i+1}^{\dagger}c_i +\bar{c}_{i+1}^{\dagger}\bar{c}_i + h.c.) \\ & + U_1(n_{i+1}n_i+\bar{n}_{i+1}\bar{n}_i) - U_2(n_i\bar{n}_i).
	\end{split}
	\end{equation}
	Here, $c_i$ ($\bar{c}_i$) is the annihilation operator for the electron (hole) at site $i$ of the upper (lower) chain. We fix the intra-chain hopping strength to be unity and $U_2, U_1>0$ corresponding to the (electron-hole) attractive and (electron-electron or hole-hole) repulsive interaction within and between chains. Note that we ignore any interchain tunneling, and the Pauli principle is explicitly incorporated in Eq.~\eqref{model} since we ignore spins as a nonessential complication for the physics of exciton condensation. Throughout this work, we also fix the filling of each chain to be $1/2$ and use the periodic boundary condition. \red{The model~\eqref{model} is closely related to the class of bilayer Hubbard models \cite{Falicov1969,Kunevs2015,Vahala2015,Kaneko2013,Rademaker2013,Rademaker2013b} or bilayer Heisenberg models \cite{Zapf2014, Sommer2001}. These classes of models feature a superfluid phase and a staggered/checkerboard insulating phase where each site is occupied by a electron of one species and a hole of the other species - an exciton. In the following calculation, we refer to the superfluid phase as bosonic liquid and the insulating phase as bosonic crystal. We note that if one extend the interaction range, the bosonic crystal is possible for lattices with less than $1/2$ filling (see the Appendix).}
	
	For $U_2=0$, each chain has an electron (hole) liquid, which is a Luttinger liquid (but this is not relevant for the physics of our interest where the focus is on the inter-chain bosonic correlations for non-zero $U_2$). To characterize the bosonic nature of the system, we compute the following correlation functions
	\begin{equation}\label{eq2}
	C^*(i',i) = \langle c_{i'}\bar{c}_{i'}\bar{c}^\dagger_{i}c^\dagger_i\rangle,~C(i',i) = \langle c_{i'}c^\dagger_{i} \rangle
	\end{equation} 
	where $C^*$ is the double-chain propagation involving moving a hole and an electron simultaneously, while $C$ is the single-chain electron propagation (hole propagation $\bar{C}$ is defined similarly). When the two chains are decoupled, i.e. $U_2=0$, $C^*(i',i)=C(i',i)\bar{C}(i',i)$ so the difference
	\begin{equation}
	b(r) = 4L^{-1}\sum_{i=1}^L \left|C^*(i+r,i) - C(i+r,i)\bar{C}(i+r,i) \right|
	\end{equation}
	indicates the bosonic correlator for the exciton propagation. The numerical coefficient $4$ is chosen so that $b(0)=1$ in the maximally coupled limit $U_2\to\infty$ where every rung is either empty or occupied by one exciton with no uncoupled electrons or holes. Indeed, in this limit, $C^*(i, i), C(i, i), \bar{C}(i, i)$ that count the onsite occupancy of excitons, electrons, and holes, all read $1/2$ due to the half-filling and charge-neutrality conditions. The quantity $b(0)$ thus indicates how strong the electron-hole bound state or equivalently the bosonic correlation is. 
	
	In Fig.~\ref{fig1}(a), we show $b(0)$ as a function of interaction strengths $U_1$ and $U_2$ in a 14-rung ladder system, i.e., 28 sites, with $b(0)>0$ for any non-zero $U_2$. For small $U_1$, the exciton becomes tightly bound for $U_2\gtrsim 3$, resulting in a vertical crossover. For large $U_1$, $b(0)$ saturates trivially because the strong repulsive $U_1$ induces independent Wigner crystals on the two chains, and any $U_2>0$ can lock these crystals with each other. We call the resultant state coherent Wigner crystal (CWC) because of the locking of the two crystals due to non-zero (albeit small) $U_2$.
	
	The bosonic correlation also has information about the mobility of the exciton bound state, which is encoded in $b=\sum_{r=1}^{L/2}b(r)$ shown in Fig.~\ref{fig1}(b). At small $U_1$ and $U_2$, the onset of long-range bosonic correlation coincides with the formation of bosonic bound states. We note that the boson in the bosonic liquid phase is actually hard-core and acts similar to fermions in 1D. Thus this description does not contradict with the Luttinger liquid description of fermions. However, naming this phase ``bosonic liquid'' emphasizes the electron-hole bound states we are interested in and provides an intuitive distinction to the bosonic crystal phase at larger $U_2$.
	
	The exciton mobility vanishes at larger $U_2$ even where $b(0)$ clearly indicates the existence of strongly bound excitons. This is because excitons in our 1D model are really hard-core bosons, which behave similarly to fermions in one dimension and tend to localize (rather than condense) under a repulsive interaction. This feature at large $U_2$ is specific to our 1D model and would not apply to 2D bilayers. To estimate the localization crossover in this large $U_2$ limit, we consider a simplified half-filled 2-rung ladder with $U_1=0$. The energy spectrum can be obtained by solving the $4\times 4$ matrix
	\begin{equation}\label{reduce}
	h= \begin{pmatrix}
	-U_2 & 1 & 1 & 0\\
	1 & 0 & 0 & 1 \\
	1 & 0 & 0 & 1 \\
	0 & 1 & 1 & -U_2
	\end{pmatrix}.
	\end{equation}
	The ground state and first excited state energies are $E_0=-(U_2+\sqrt{16+U_2^2})/2$ and $E_1=-U_2$, \red{separated from the other two levels by $\sim U_2$. In the limit of large $U_2$, the lowest-energy configurations are those with electrons and holes occupying the same rung. As such, the transition from one low-energy configuration to another can be thought of as the hopping of an exciton whose strength induces a splitting between the two lowest-energy levels}
	\begin{equation}
	t^* = (E_1 - E_0)/2 = \left(\sqrt{16+U_2^2}-U_2\right)/4.
	\end{equation}
	The bosonic liquid crosses over to the solid when $t^*\propto U_1$ [we demonstrate $t^*=U_1/4$ in Fig.~\ref{fig1}(b)]. For $U_2\to 0$, the crossover happens at $U_1=\mathcal{O}(1)$ while in the limit $U_2\to \infty$, $U_1\propto 1/U_2$. These analytical conditions for the transitions to the solid phase (for $U_1 \gg 1$ and $U_2\gg 1$) agree with our numerical results, allowing us to draw a putative line between the bosonic liquid and the bosonic solid. We believe this transition to be an effective first-order solid-liquid transition for the exciton system that is adiabitically connected to the Luttinger liquid - charge density wave transition for $U_2=0$ \cite{Giamarchi2007,Florian2022}.
	
	To confirm our statement that the bosonic liquid can crystalline under repulsive interaction, we directly compute the crystalline modulation in the inter-chain density-density correlation 
	\begin{equation}\label{eq6}
	\eta = \frac{2}{L(\lfloor L/4 \rfloor+1)}\sum_{i=1}^L \sum_{j=0}^{L/2} (-1)^j\langle \bar{n}_i n_{i+j} \rangle.
	\end{equation}
	We note that the alternating sign is compatible with the half-filled crystal, and the normalization guarantees $\eta=1$ for the maximally crystallized state. Figure~\ref{fig1}(c) clearly shows that the vanishing of exciton mobility almost coincides with the formation of bosonic crystals. The very slight mismatch can be explained by Eq.~\eqref{reduce}. Accordingly, the total energy of two excitons (four fermions) $E_\text{pair} = -2U_2 > E_1 + E_0$. This means, even without the long-range $U_1$, two excitons have an exchange-like repulsive interaction in the same order as the kinetic energy. Therefore, the strongly-bound bosonic liquid has non-zero crystalline order even though the electrons and holes have no intrinsic repulsive interaction. This explains why the crystalline order emerges earlier than the vanishing of the bosonic liquid phase. The bosonic solid-liquid phase boundary, together with the weakly (small $U_2$) to strongly (large $U_2$) bound states, partition the ($U_1, U_2$) parameter space into four regimes. The bosonic liquid is separated into the weakly and strongly bound BEC -- the so-called BCS-BEC crossover. Similarly, the bosonic crystal phase can be subdivided into the CWC and dipolar crystal (DC). CWC and DC are not different phases and differ only quantitatively, depending on whether the state is induced by introducing weak $U_2$ between the two Wigner crystals or $U_1$ on the liquid phase of strongly bound excitons.
	
\begin{figure}
		\centering
		\includegraphics[width=0.37\textwidth]{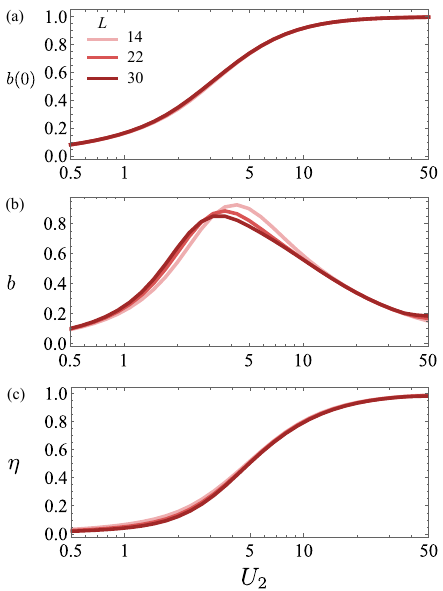}
		\caption{Finite-size scaling analysis of (a) $b(0)$, (b) $b$, and (c) $\eta$ at $U_1=0.5$. $L$ is the number of rungs in the ladder model. \label{finitesize}}
	\end{figure}

	One remaining question is whether the weakly bound exciton survives or is replaced by electron/hole plasma in the thermodynamic limit. In Fig.~\ref{finitesize}, we perform the finite-size scaling analysis with respect to Fig.~\ref{fig1}. The results converge with system sizes, which supports the claim that excitons always form for any attractive interaction $U_2>0$, and there is no Mott transition for any finite $U_2$. We find that the calculated bosonic correlation decreases continuously with decreasing $U_2$ without vanishing, thus arguing for the weak-coupling excitonic  BCS state to be stable down to arbitrarily small $U_2$ within our numerical accuracy.		
	
	\section{Disordered phase diagram } 

	\begin{figure*}
	\centering
	\includegraphics[width=\textwidth]{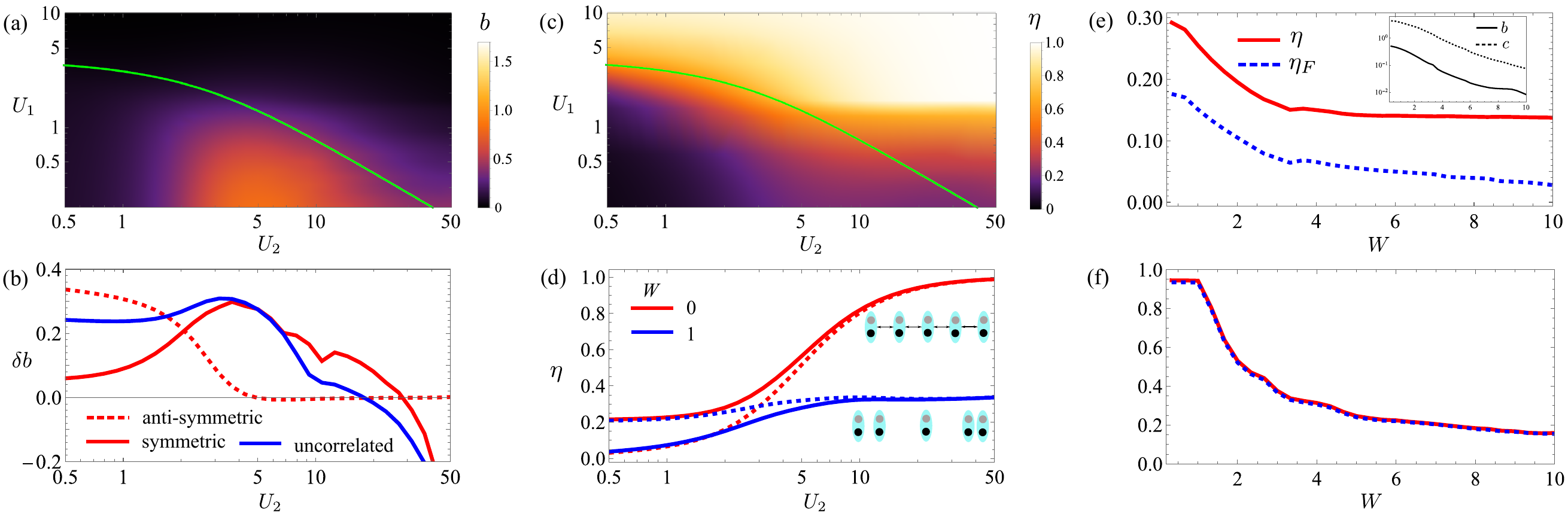}
	\caption{ (a) Bosonic long-range correlation under uncorrelated disorder and fixed $W=1$, the crossover line is carried over from the pristine system. (b) Relative deviation induced by three disorder models having the same $W=1$ along the line $U_1=0.5$. (c) The crystalline order of the disordered system is similar to (a). (d) Comparison between $\eta$ (solid) and $\eta_F$ (dashed), showing the inter-chain synchronization happens at the same $U_2$. (e, f) $\eta$ and $\eta_F$ for $U_1=0.5, U_2=2$(e) and $U_1=2, U_2=10$ (f). The inset in (e) shows that the bosonic and fermionic correlations decay concomitantly with $W$.   \label{fig3}}
\end{figure*}  		
	
	We study the robustness of the phase diagram under random disorder, i.e., $H=H_0+\sum_i W_in_i + \bar{W}_i\bar{n}_i$. The disorder value at each site is drawn independently from a uniform distribution $(-W, W)$. 
	
	In Fig.~\ref{fig3}(a) and (c), we reproduce the phase diagram under uncorrelated noise with amplitude $W=1$. This noise dampens the long-range bosonic correlation [Fig.~\ref{fig3}(a)] and generates a peculiar regime around the crossover line for $U_1<2$ where the bosons are nearly immobile ($b\sim 0$) but do not crystallize either ($\eta<1$). Our current model assumes no correlation between the disorder on the upper and lower chains. We can, however, gain some insight from studying the problem using two complementary limits, i.e., the disorders on the two chains are (i) anti-symmetric ($W_i=-\bar{W}_i$) or (ii) symmetric ($W_i=\bar{W}_i$).
	
	In Fig.~\ref{fig3}(b), we show the relative deviation on the long-range bosonic correlation $\delta b = (b_0-b_W)/b_0$ where $b_W$ is computed at disorder strength $W$ (and fixed $U_1=0.5$). Despite having the same disorder amplitude $W=1$, the three disorder models behave differently. The anti-symmetric disorder is relevant for the weak-coupling $U_2$ but negated by sufficiently large $U_2$. The anti-symmetry of the disorder model tends to localize electrons and holes at different rungs (a local potential minimum in one chain corresponds to a local maximum in the adjacent), directly competing with the effect of the inter-chain attraction that prefers both an electron and a hole on one rung. For stronger $U_2$, the elementary particles are point-like composite bosons, and the net potential felt by this object is vanishing, resulting in robustness against disorder.
	
	By contrast, the system is more susceptible to symmetric disorder in the strong-coupling $U_2$ limit. In this limit, the net potential the boson feels is non-zero and $\sim W$. At the same time, the effective hopping is suppressed by $U_2^{-1}$, resulting in the enhanced sensitivity to symmetric disorder for large $U_2$. Returning to our original uncorrelated disorder model, it can be decomposed into anti-symmetric and symmetric components. Therefore, for small $U_2$, disorder affects the system by separately localizing electrons and holes (similar to anti-symmetric disorder). For large $U_2$, the main effect is the localization of tightly-bound bosons (electrons and holes localized at adjacent sites similar to symmetric disorder). This observation explains the irregular localized regime in Fig.~\ref{fig3}(b). This is the Bose glass regime induced by disorder (for $U_2>5$ in Fig.~\ref{fig3}b), where bosons are localized.
	
	In the presence of uncorrelated disorder, the two chains are not equivalent, so we introduce two additional order parameters, which are the intra-chain version of $b$ and $\eta$
	\begin{equation}
	\begin{split}\label{eq7}
	c =  4L^{-1}\sum_{r=1}^{L/2}\sum_{i=1}^L \left|C(i+r,i)\bar{C}(i+r,i) \right|\\
	\eta_F = \frac{2}{L(\lfloor L/4 \rfloor+1)}\sum_{i=1}^L \sum_{j=0}^{L/2} (-1)^j\langle n_i n_{i+j} \rangle.
	\end{split}
	\end{equation}
	In Fig.~\ref{fig3}(d), we compare $\eta$ and $\eta_F$ in the pristine and disordered cases at fixed $U_1=0.5$.  The convergence of these two quantities indicates the synchronization between the two chains, which happens around the same $U_2$ for both cases. However, in the pristine case, $\eta$ and $\eta_F$ are both saturated, but not in the disordered case. The corresponding boson localization landscapes are pictorially shown in Fig.~\ref{fig3}(d) with the long-range order present (absent) in the pristine (disordered) system. 
	
	The previous results are all given for $W=1$. We now study whether stronger disorder can destroy inter-chain coherence. In Fig.~\ref{fig3}(e), we show the effect of disorder on the bosonic liquid phase ($U_1=0.5, U_2=2$). It is clear from the inset that $b$ is suppressed exponentially by $W$, but the intra-chain counterpart $c$ also decays just as fast. This shows that the bosonic correlation is destroyed at the same time as the underlying Fermi surface, and there cannot exist any decoherent Luttinger liquid phase by introducing disorder. Additionally, $\eta$ and $\eta_F$ stay distinct for any $W$, clarifying that the large-$W$ state is two independent Anderson insulators. On the other hand, starting with the bosonic crystal ($U_1=2, U_2=10$), $\eta$ and $\eta_F$ decrease but remain identical, showing a transition from bosonic crystal to bosonic glass. Thus, as $W$ increases, the bosonic liquid becomes decoherent Anderson insulators, while the bosonic crystal loses the long-range order and becomes bosonic glass. In the limit $W\to\infty$, the entire parametric phase is trivially decoupled electron/hole Anderson insulators.   
	
	\section{Effect of finite temperature}		

	\begin{figure}
		\centering
		\includegraphics[width=0.48\textwidth]{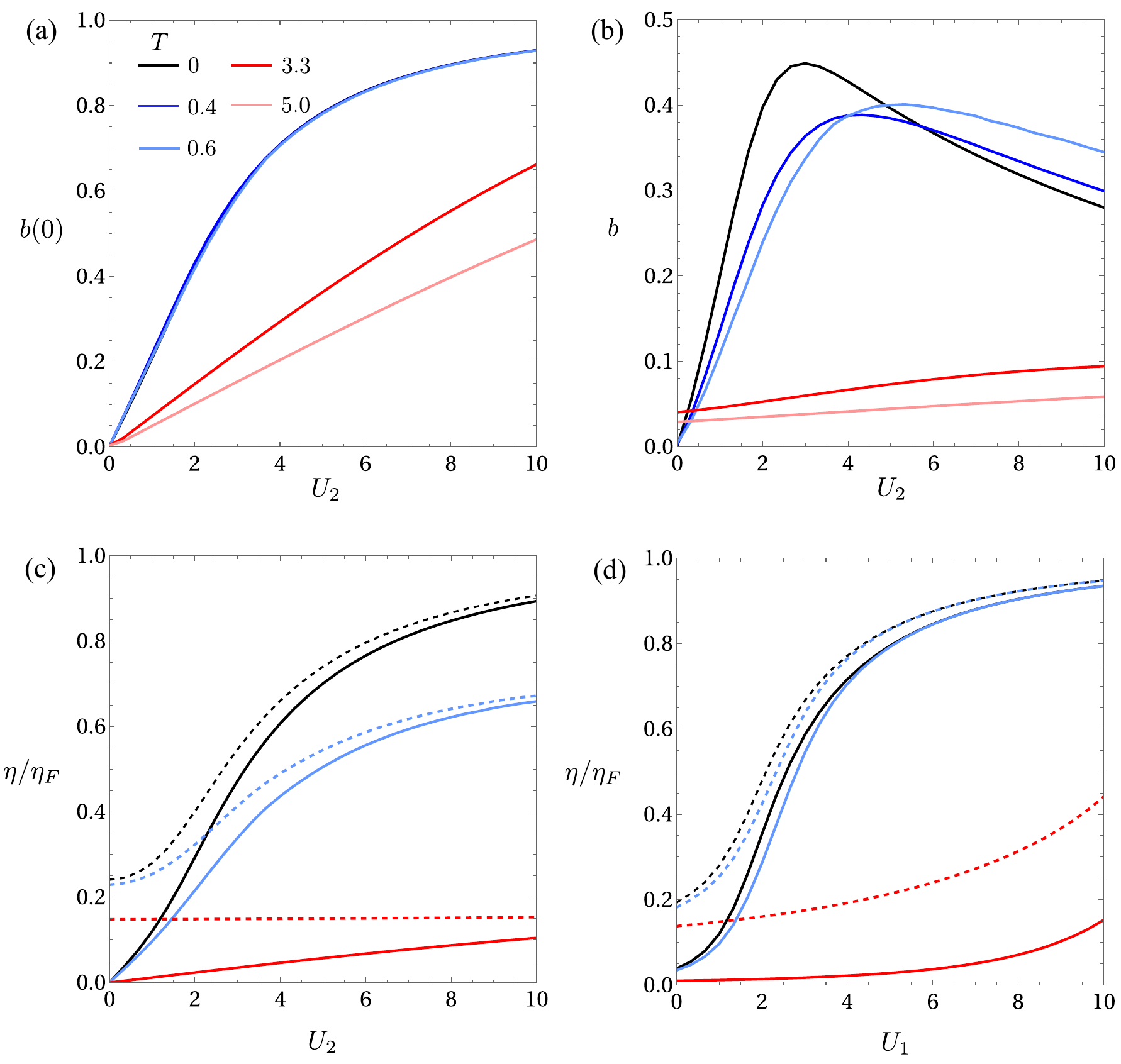}
		\caption{ (a) Onsite bosonic correlation, (b) long-range bosonic correlation. (c-d) Excitonic (solid lines) and fermionic (dashed lines) crystalline order with respect to $U_2$ and $U_1$. Blue and light blue colors denote low temperatures (less than exciton binding energy), while red and light red denote high temperatures (more than exciton energy).   \label{hightemp}}
	\end{figure}  	

	This section studies the thermal melting and crossover to classical phases at low ($T\ll 1$) and high ($T\sim U_1, U_2$) temperatures. For low temperatures, we compute the first 29 excited states and take the thermal averages of observables
	\begin{equation}
		\langle O \rangle_\beta = \frac{\text{Tr}\left(e^{-\beta H}O\right)}{\text{Tr}\left(e^{-\beta H} \right)} \approx \frac{ \sum_{i=0}^{29} e^{-\beta E_n} \bra{\psi_n} O \ket{\psi_n}}{\sum_{n=0}^{29} e^{-\beta E_n}},
	\end{equation}
	which enter the correlation functions ~\eqref{eq2}, \eqref{eq6} and \eqref{eq7} with $O$ being the corresponding operator. We note that the chemical potential does not appear because we explicitly impose particle number conservation on each chain when computing excited states.  
	
	For high temperatures $\beta \approx 0$, we use the stochastic sampling method combined with imaginary time evolution. We first set the initial state by assigning the particles (fixed particle numbers on each chain) to a set of randomly chosen single-particle levels ($U_1=U_2=0$) so that the initial state ensemble consists of orthonormal Slater states. We then evolve the state using the TDVP method and compute the observable expectation value at imaginary time $\tau = \beta/2$. The thermal expectation value is obtained by averaging over the ensemble of initial states as
	\begin{equation}
		\langle O \rangle_\beta = \frac{\text{Tr}\left(e^{-\beta H}O\right)}{\text{Tr}\left(e^{-\beta H} \right)} \approx  \frac{\sum_{i=1}^{N} \bra{\psi_i} e^{-\beta H/2}  O e^{-\beta H/2} \ket{\psi_i}}{\sum_{i=1}^{N} \bra{\psi_i} e^{-\beta H}\ket{\psi_i}}
	\end{equation}
	which we choose $N=200$. We have numerically tested the convergence for $N=200$.
	
	For temperatures lower than exciton binding energy (blue and light blue lines), the onsite correlation $b(0)$ is unaffected (Fig.~\ref{hightemp}(a)), but the long-range bosonic correlation decreases (increases) compared to the zero-temperature bosonic liquid (crystal) phase (Fig.~\ref{hightemp}(b)). This is consistent with the thermal melting of the bosonic crystal into a bosonic liquid phase, which is further supported by Fig.~\ref{hightemp}(c) with $\eta$ approaching $\eta_F$ but not saturating as $U_2$ increases. On the other hand, nearest-neighbor $U_1$ (as compared to the onsite $U_2$) can recover the crystalline order and increase the melting temperature, as can be seen in Fig.~\ref{hightemp}(d). Bosonic phases are only destroyed when the temperature exceeds the exciton binding energy, replaced by classical electron and hole plasma. Thus, at finite $T$, there can indeed be an electron-hole liquid phase, particularly for small $U_2$ where the exciton binding is weak.	
	
	\section{Conclusion}
	\begin{figure}
	\centering
	\includegraphics[width = 0.3\textwidth]{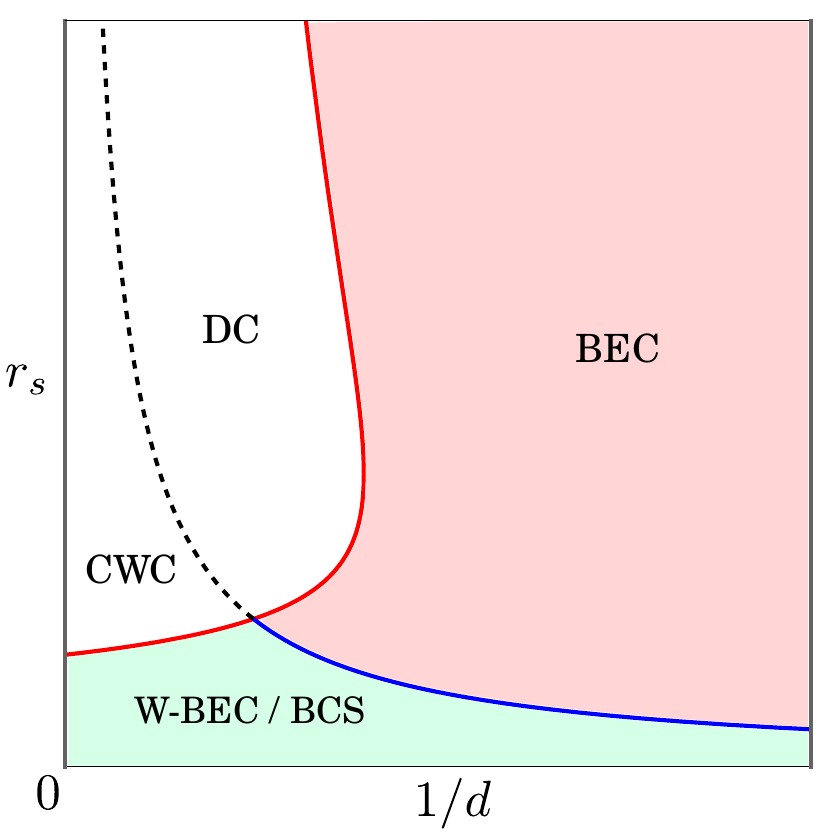}
	\caption{Schematic phase diagram of indirect excitons in 2D bilayers. The red line given by Eq.~\eqref{sketch} separates the crystalline and liquid phases, and the blue line given by $r_s\sim d$ separates the strongly bound exciton BEC phase from the weakly-bound BEC (W-BEC) or the BCS phase. The extension of the blue line into the crystalline phase (dashed lines) indicates the dipolar crystal (DC) and the coherent Wigner crystal (CWC) phases. There is no electron-hole plasma phase at $T=0$. \label{diagram}}
	\end{figure} 	
	
	We have calculated the quantum phase diagram of indirect excitons formed between two oppositely charged spatially separated  1D chains. This setup lets us directly tune the binding energy and measure various correlation functions. At zero temperature, excitons always form, and the entire parameter space is partitioned into the bosonic liquid and bosonic crystal phases. Upon introducing a small disorder, the bosonic characteristic remains unchanged, with the bosonic crystal phases transitioning into a bosonic Anderson insulator. At finite $T$, there can indeed be an electron-hole liquid phase, particularly for small $U_2$ where the exciton binding is weak.
	
	Based on our 1D tight-binding 2-channel ladder results, we propose some features of the continuum excitons, potentially realized in 2D bilayer systems. The kinetic and intra-layer repulsive interaction is governed by the electron/hole density so that $t\sim 1/r_s^2$ and $U_1\sim 1/r_s$. Here we use the convention $r_s=a$ with $a$ being the average inter-particle spacing within one layer normalized by the Bohr radius. $U_2$ in our model becomes $1/d$ with dimensionless $d$ the inter-layer separation also scaled by the Bohr radius. There are two differences with our tight-binding model. First, the exciton kinetic energy (or that of the electron-hole pair center of mass) remains finite for strong $U_2$ and only depends on $r_s$. Additionally, the interaction in the strong-$U_2$ limit becomes a dipolar $1/r^3$ interaction. Note that the strong $U_2$ limit is different in 2D bilayers compared with 1D since hard-core bosons are no longer equivalent to fermions. We estimate a phase diagram as shown in Fig.~\ref{diagram} where the qualitative phase boundary is mostly likely crossover in 1D and true phase transition in 2D. The solid-liquid phase boundary is sketched as follows
	\begin{equation}\label{sketch}
	\frac{1}{r_s} - \frac{1}{\sqrt{r_s^2+d^2}} = \frac{C}{r_s^2}
	\end{equation}
	with the LHS being the repulsive interaction energy, RHS the kinetic energy (electrons/holes and excitons), and $C$ a constant. The exciton description apparently prevails when $d\ll a$ or $r_s \gg d$. We note that the asymptotic Eq.~\eqref{sketch} yields $r_s \sim d^2 \gg d$ in the large-$d$ limit, so similar to the tight-binding model, the solid phase may qualitatively be divided into the coherent Wigner crystal and dipolar crystal phases depending on the value of $r_s/d$. In Appendix A, we perform a numerical simulation with attractive long-range interaction to capture the dipole physics. Remarkably, a part of this proposed phase diagram is realized in our dipolar 1D tight-binding model, with the main difference being the large $U_2$ regime.\\
	
	\textit{Note added:} The preprint \cite{Zeng2023} claims to have observed experimentally the coherent exciton crystal phase that we find in our simulations.
	
	\begin{acknowledgments}
		The authors are grateful for helpful discussions and communications with Professors Jay Deep Sau, Bert Halperin, Peter Littlewood, Kin Fai Mak, Fengcheng Wu, Jim Eisenstein, and Yang-Zhu Chou. This work is supported by Laboratory for Physical Sciences. The authors acknowledge the University of Maryland supercomputing resources (\href{https://hpcc.umd.edu}{https://hpcc.umd.edu}) made available for conducting the research reported in this paper.
	\end{acknowledgments}
	
	\begin{figure*}[h!]
	\centering
	\includegraphics[width=\textwidth]{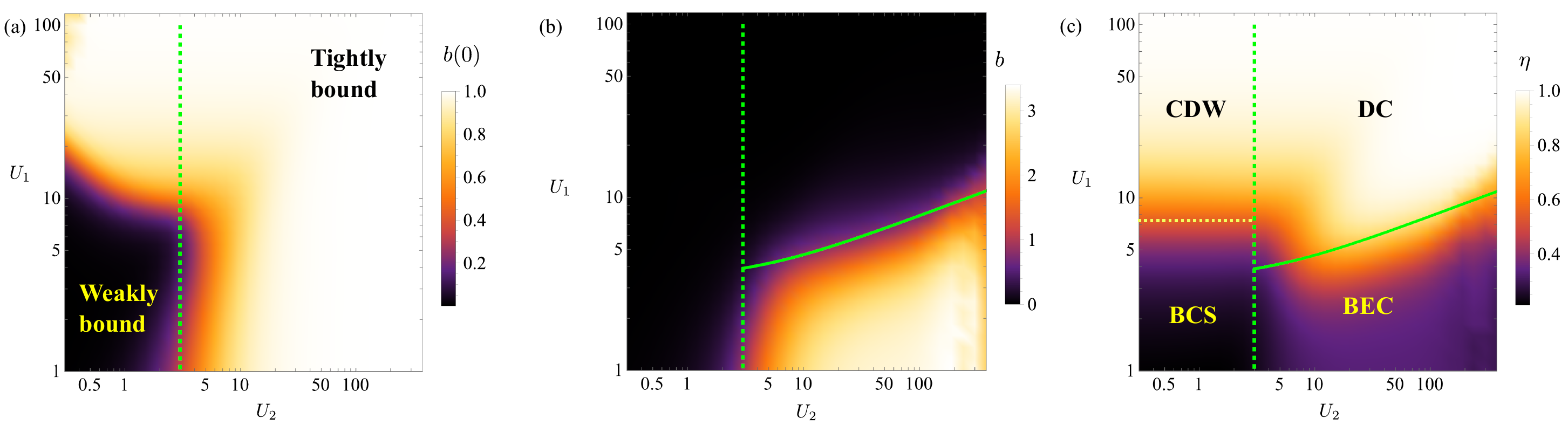}
	\caption{($U_1, U_2$) phase diagram for a 14-rung ladder system with long-range $\gamma=1.2$ inter-chain attractive and intra-chain repulsive interactions. (a) Onsite bosonic correlation $b(0)$. (b) Long-range bosonic correlation. (c) Crystalline order parameter from the bosonic density-density correlation. The dashed line given by $U_2=3$ separates the weakly bound $(b(0)<0.5)$ and tightly bound $(b(0)>0.5)$ exciton regimes at weak $U_1$. The solid green line given by Eq.~\eqref{line} with $C=1.2$ separates the bosonic crystal and bosonic liquid. Note that the large $U_2$ (and small $U_1$) regime for the dipolar model here is qualitatively different from that in the adjacent attractive interaction model of Fig.~\ref{fig1} in the main text.\label{dipole}}
\end{figure*}

\begin{figure*}[h!]
	\centering
	\includegraphics[width=\textwidth]{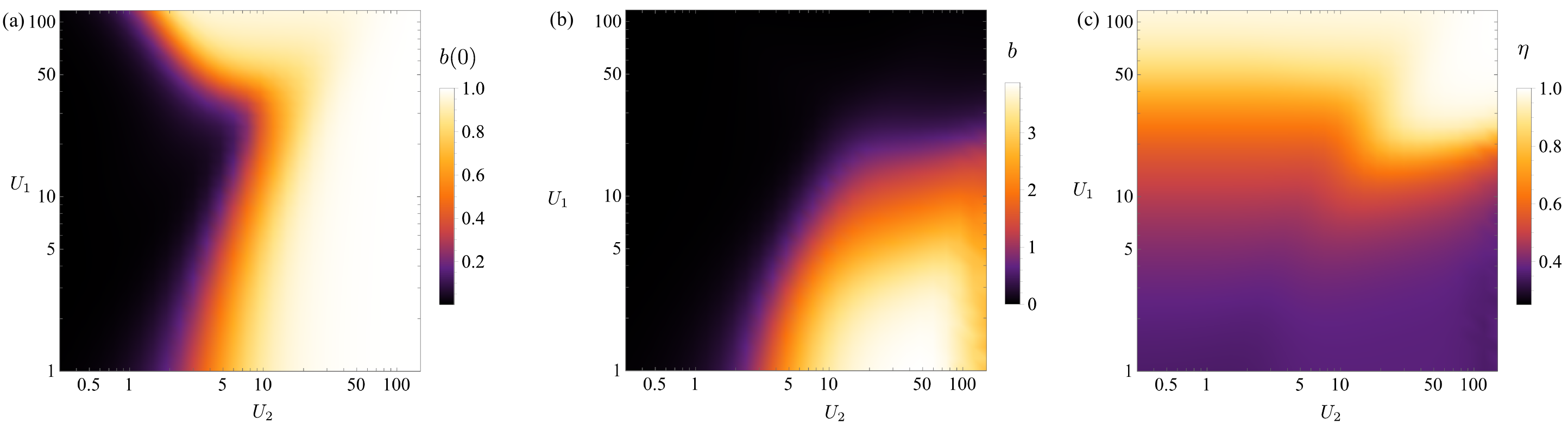}
	\caption{Same as Fig.~\ref{dipole} but with 15 sites per chain and the filling of each chain being $1/3$. The normalization coefficients is $9/2$ for (a-b). For (c), the crystalline order parameter becomes $\eta = \frac{3}{L(\lfloor L/6 \rfloor+1)}\sum_{i=1}^L \sum_{j=0}^{L/2} \cos\frac{2\pi j}{3}\langle \bar{n}_i n_{i+j} \rangle$. The phase diagram is qualitatively similar to Fig.~\ref{dipole} but the extended phases (BCS and BEC) expands along $U_1$ axis. \label{dipole2}}
\end{figure*}	
	
	\appendix
		\section{Long-range attractive potential}

	\red{We modify the interaction potential into
\begin{equation}
	\begin{split}
		& H_I = -\sum_i U_2 n_i\bar{n}_i + \sum_{i<j} (n_i, \bar{n}_i) U_{i,j} (n_j,\bar{n}_j)^T,\\
		& U_{i,i+\delta} = \begin{pmatrix}
			\frac{U_1}{\delta^\gamma} && - \frac{U_1U_2}{(U_1^{2/\gamma}+\delta^2 U_2^{2/\gamma})^{\gamma/2}} \\
			- \frac{U_1U_2}{(U_1^{2/\gamma}+\delta ^2 U_2^{2/\gamma})^{\gamma/2}} && \frac{U_1}{\delta^\gamma}		
		\end{pmatrix} 
	\end{split},
\end{equation}
assuming that the elementary interaction $\sim 1/r^\gamma$ so that the intra-chain repulsive interaction scales as $a^{-\gamma}$ with $a$ being the site spacing and the inter-chain attractive interaction scales as $(d^2+a^2)^{-\gamma/2}$ with $d$ being the separation between the two chains.} In Fig.~\ref{dipole}, we show an equivalence of Fig.~\ref{fig1} in the main text. The main difference with the system in the main text is that the long-range interaction modifies the solid-liquid phase boundary into
\begin{equation}\label{line}
	\sqrt{16+U_2^2} - U_2 = C U_1\left(1 - \frac{U_2}{(U_1^{2/\gamma} + U_2^{2/\gamma})^{\gamma/2}}\right),
\end{equation}
where the RHS is the nearest-neighbor dipole interaction. We can see that in the limit $U_2\to \infty$, the hopping is renormalized to $t^*\sim 1/U_2$, while interaction now becomes dipole-dipole interaction and reads $U_1^3/U_2^2 \ll t^*$. As a result, the large-$U_2$ regime is dominated by the BEC phase, unlike the crystal phase in Fig.~1 where large $U_2$ converts the effectively hard core bosonic system to a crystalline fermionic system for $U_2 \gg 1$.

\red{Due to the long-range nature of the interaction, insulating phase is possible even for less than half-filled lattice. The constraint on the filling fraction now reduces to $1/n$ for 1D lattices or some rational numbers for higher-dimensional lattices depending on the lattice symmetry. In Fig.~\ref{dipole2}, we redo the calculation of Fig.~\ref{dipole}} with $1/3$ filling, i.e. $5$ electrons and holes occupying a 15-rung ladder. Qualitatively, the phase diagram is similar to the half-filled case but quantitatively, the liquid phase expand. This can be understood of the scaling of the dipolar interaction $1/r^3$ versus that of the kinetic energy $1/r^2$, meaning that the system favors the extended phase at low filling to minimize the kinetic energy.

	\bibliographystyle{apsrev4-2}
\bibliography{reference}
\end{document}